\documentclass[a4paper,11pt]{article}
\pdfoutput=1 

\usepackage{jcappub} 
\usepackage{aas_macros} 
\usepackage[T1]{fontenc} 
\usepackage{amsmath}
\usepackage{subcaption}

\title{\boldmath Chameleon Screening in Cosmic Voids}

\author[a]{Andrius Tamosiunas,}
\author[a]{Chad Briddon,}
\author[a]{Clare Burrage,}
\author[a]{Alan Cutforth,}
\author[a]{Adam Moss,}
\author[a]{and Thomas Vincent}

\affiliation[a]{School of Physics and Astronomy, University of Nottingham, Nottingham, NG7 2RD,\\ United Kingdom}

\emailAdd{andrius.tamosiunas@nottingham.ac.uk}
\emailAdd{chad.briddon@nottingham.ac.uk}
\emailAdd{clare.burrage@nottingham.ac.uk}
\emailAdd{ppyalcu@exmail.nottingham.ac.uk}
\emailAdd{adam.moss@nottingham.ac.uk}
\emailAdd{ppytav@exmail.nottingham.ac.uk}

\abstract{A key goal in cosmology in the upcoming decade will be to form a better understanding of the accelerated expansion of the Universe. Upcoming surveys, such as the Vera C. Rubin Observatory's 10-year Legacy Survey of Space and Time (LSST), Euclid and the Square Killometer Array (SKA) will deliver key datasets required to tackle this and other puzzles in contemporary cosmology. With this data, constraints of unprecedented power will be put on different models of dark energy and modified gravity. In this context it is crucial to understand how screening mechanisms, which hide the deviations of these theories from the predictions of general relativity in local experiments, affect structure formation. In this work we approach this problem by using a combination of analytic and numerical methods to describe chameleon screening in the context of cosmic voids. We apply a finite element method (FEM) code, SELCIE, to solve the chameleon equation of motion for a number of void profiles derived from observational data and simulations. The obtained results indicate a complex relationship between the properties of cosmic voids and the size of the chameleon acceleration of a test particle. We find that the fifth force on a test particle in a void is primarily related to the depth and the inner density gradient of the void. For realistic void profiles, the obtained chameleon-to-Newtonian acceleration ratios range between $a_{\phi}/a_{\rm Newt} \approx 10^{-6} - 10^{-5}$. However, it should be noted that in unusually deep voids with large inner density gradients, the acceleration ratios can be significantly higher. Similarly, other chameleon models, such as $f(R)$ Hu-Sawicki theory allow for significantly higher acceleration ratios. Given these results, we also discuss the optimal density profiles for detecting the fifth force in the upcoming observational surveys.}

\begin{document}
\maketitle
\flushbottom
\raggedbottom

\section{Introduction}

We are living in the era of the accelerated expansion of the Universe, as illustrated by the wealth of data from type Ia supernovae, baryon acoustic oscillations (BAO), galaxy clusters and other observational probes \cite{Riess1998, Sahni2000,Peebles2003, Lapuente2010}. Arguably the simplest and the most successful explanation for the accelerating expansion as of yet is the cosmological constant. Nonetheless, the discrepancy between the value of the cosmological constant determined in cosmology and that calculated in quantum field theory could hint towards a more complicated model for the accelerated expansion. Similarly, contemporary problems in cosmology, such as the $H_{0}$ tension, could be hinting towards new physics. In this context, modifying general relativity offers an alternative explanation to some of the puzzles outlined above \cite{Joyce2016, Clifton2012, Di_Valentino2021, Braglia2021, Benisty2021}. 

A subset of theories of special interest are scalar field models that possess a screening mechanism. These are theories that turn off modified gravity effects in and around high density regions, hence satisfying the tight Solar System observational constraints, while still possibly allowing interesting effects on cosmological scales. Well known examples of screening mechanisms include the chameleon, the symmetron and the Vainshtein models \cite{Vainshtein1972, Khoury2004, Hinterbichler2011, Babichev2013}. Despite the elusive nature of screened fifth forces mediated by these types of scalar fields, there exists a number of techniques to detect them in laboratory and cosmology tests. In particular, extensive cosmological and laboratory constrains have been put on chameleon screening \cite{Terukina2014, Wilcox2015, Burrage2018, Sabulsky2019, Tamosiunas2021}. Similarly, techniques for constraining symmetron and Vainshtein screening are well described in the literature as well \cite{Llinares2014, Burrage2016, Jimenez2016, Dima2018, Hammami2017, Llinares2019}. 

In the context of cosmological tests of gravity, a system of special interest is that of cosmic voids. Voids are the largest underdensities in the Universe and their properties are inherently linked to the physics of large scale structure. Given these properties, cosmic voids make ideal cosmological laboratories for testing screened gravity theories. Using voids as a testing ground for models of modified gravity has been explored extensively in the literature (e.g. see Ref.~\cite{Cai2015, Perico2019, Padilla2015, Davies2019}). Nonetheless, many unanswered questions regarding the properties of cosmic voids remain to be addressed. Specifically, an interesting question to ask is how does the fifth force depend on the density distribution in cosmic voids? Previous work has been done to deduce the magnitude of the fifth force in voids; however, these analyses have been performed primarily using a top hat/step function density models for the voids (e.g. see Ref.~\cite{Clampitt2013}). Recent simulations and observational studies indicate that void density distributions are significantly more complicated than that of a step function. Hence, one might ask what is the relation between different void density distributions and the magnitude of the fifth force? As an extension of our previous numerical work (Ref.~\cite{Tamosiunas2021}), here we attempt to answer this question by numerically solving the chameleon equation of motion (EOM) for different void density profiles.  

In the following sections we introduce the model at hand along with the numerical methods used to solve the EOM. Specifically, section \ref{chameleon_gravity} introduces the specific chameleon model that we will investigate. Section \ref{finite_element_method} briefly describes the SELCIE code used to solve the chameleon EOM. Section \ref{cosmic_voids} introduces some basic properties of voids, with a particular emphasis on different density profiles used in our analysis. The analytic and the numerical results are described in sections \ref{analytic_investigation} and \ref{numerical_investigation}. Appendix~\ref{appendix A} shows a comparison between the results obtained in this work and the corresponding results for certain $f(R)$ theories. In this work we use natural units ($c = \hbar =1$) and a mostly positive metric signature $\eta_{\mu \nu} = \mathrm{diag}(-1,1,1,1)$. We denote the reduced Planck mass as $M_{\rm Pl} = (8 \pi G)^{-1}$, with $G$ as the Newton's constant. In all the shown figures we used rescaled radial units, such that $\hat{r} = r/L$, with $L = 2 \times 31.68$ Mpc/$h$ (twice the typical void size as described in Ref.~\cite{Chantavat2017}). Similarly, the densities are rescaled by the critical density of the Universe $\hat{\rho} = \rho/\rho_{\mathrm{c}}$ with $\rho_{\mathrm{c}}=3 H^{2}/8 \pi G=1.8788 \times 10^{-26} h^{2} \mathrm{~kg} \mathrm{~m}^{-3}$ with $h = 0.673$.

\section{Chameleon Gravity}
\label{chameleon_gravity}

A non-minimally coupled scalar-tensor theory is described by the following action:

\begin{equation}
S=\int \mathrm{d} x^{4} \sqrt{-g}\left(\frac{M_{\mathrm{Pl}}^{2}}{2} R-\frac{1}{2} \nabla_{\mu} \phi \nabla^{\mu} \phi-V(\phi)\right)-\int \mathrm{d} x^{4} \mathcal{L}_{m}\left(\varphi_{m}^{(i)}, \tilde{g}_{\mu \nu}^{(i)}\right),
\end{equation}

\noindent with $\phi$ as the scalar field, $V(\phi)$ as the potential, $\mathcal{L}_{m}$ as the matter Lagrangian, $\varphi_{m}^{(i)}$ as the matter fields and $\tilde{g}_{\mu \nu}^{(i)}$ as the Jordan frame metric. The superscript \textit{i} here refers to the i-th matter species. The Jordan frame metric can be related to the Einstein frame metric via the following relation:

\begin{equation}
\tilde{g}_{\mu \nu}^{(i)}=A_{i}^{2}(\phi) g_{\mu \nu}.
\end{equation}

\noindent Chameleon gravity refers to a particular choice of $A_{i}$ and $V(\phi)$, e.g.

\begin{equation}
    A_{i}(\phi) = e^{\beta_{i} \phi/M_{Pl}},
\end{equation}

\noindent with $\beta_{i}$ term here as the coupling constant for the i-th matter species and 

\begin{equation}
V(\phi)=\Lambda^{4}\left(1+\frac{\Lambda^{n}}{\phi^{n}}\right),
\label{potential}
\end{equation}

\noindent with an energy scale $\Lambda$ (note that other choices of the potential that lead to chameleon screening are possible). Assuming a universal coupling to matter, the EOM is given by:

\begin{equation}
\nabla^{2} \phi=-\frac{n \Lambda^{n+4}}{\phi^{n+1}}+\frac{\rho}{M}.
\label{eq:EOM}
\end{equation}

\noindent Note that here we have also assumed a time-independent scalar field and defined $M \equiv M_{\mathrm{Pl}} / \beta$, where we dropped the index for $\beta$ under the assumption of a universal coupling constant.

\begin{figure}
  \centering
    \includegraphics[width=0.85\textwidth]{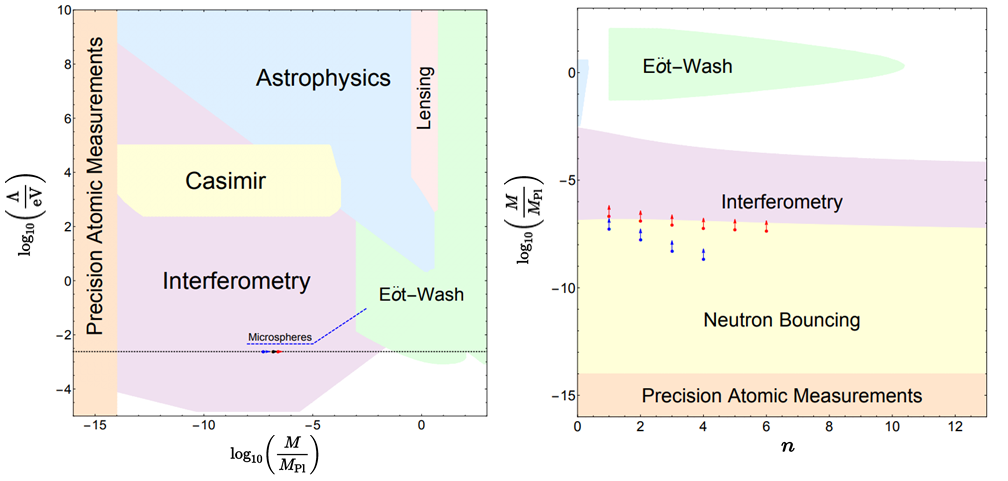}
    \caption{Observational and laboratory constraints of chameleon gravity. The regions excluded by each specific test are indicated in the figure. The figure on the left shows the constraints on $\Lambda$ and $M$, for $n = 1$. The figure on the right shows the corresponding constraints for $n$ and $M$ for $\Lambda = 2.4$ meV. The region labelled astrophysics contains the bounds from both Cepheid and rotation curve tests. The dashed line indicates the dark energy scale $\Lambda = 2.4$ meV. The black, red, and blue arrows in both figures show the lower bound on $M$ coming from neutron bouncing and interferometry experiments. In this work we focus on the white regions of the parameter space. Figure adapted from \cite{Burrage2018}. }
    \label{chameleon_constraints}
\end{figure}

For the above potential, Eq.~\ref{potential}, the different chameleon models correspond to different choices of the three parameters: $n$, $\Lambda$, $M$. In this work we primarily focus on the models that are still allowed by the current observational constraints (see Fig.~\ref{chameleon_constraints}). Specifically, we investigate the parameter space regions still allowed by the observational constraints to determine which region would produce the highest values for the chameleon acceleration. In accordance with our previous work in Ref.~\cite{Tamosiunas2021}, we will find that in order to maximise the chameleon acceleration, the smallest allowed value of $M$ is needed. Similarly, the smallest allowed value of $n$ is required to maximise the chameleon acceleration. Lastly, we want $\Lambda$ to be as large as possible. Putting all these requirements together the parameter values that maximise the fifth force, while still being in the allowed parameter space, are: $n = 1$, $M = 10^{-14} \times M_{\rm Pl}$ and $\Lambda = 10^{-4}$ eV.

Here we should also note that there are other chameleon models often studied in the literature, such as the $f(R)$ Hu-Sawicki model (e.g. Ref.~\cite{Hu-Sawicki2007}), which corresponds to a different region of the parameter space shown in Fig~\ref{chameleon_constraints}. Studying such chameleon models in detail is left for future work, however, an illustrative comparison between our results and the corresponding results in the Hu-Sawicki model is given in appendix \ref{appendix A}. 

The chameleon EOM is non-linear and finding exact analytic solutions for complex non-symmetric density distributions is generally not possible. Hence, to find solutions we will apply the numerical technique known as FEM. FEM allows finding solutions to non-linear equations for arbitrary density distributions. Specifically, in order to solve the EOM, we will employ the software package SELCIE, which is described further in the upcoming section.

\section{The SELCIE Solver}
\label{finite_element_method}

FEM refers to a widely used numerical technique for solving linear and non-linear differential equations in one, two or three spatial dimensions. FEM has now existed for over 80 years and its range of applications varies widely from fluid mechanics and engineering to meteorology and physics \cite{Turner1956, Liu2021, Hrennikoff2021}. Recently FEM has also been applied extensively in the context of modified gravity \cite{Braden2021, Burrage2021, Briddon2021, Tamosiunas2021}. 

FEM subdivides the problem domain into smaller subdomains referred to as finite elements. Specifically, the domain is segmented into cells, whose boundaries are defined by their vertices. The value of the field inside each cell is then approximated by a piecewise polynomial function that matches the field values at each of the cell’s vertices. For non-linear second order differential equations one can use Green's theorem and Taylor expand the non-linear term. The static solution can then be obtained at each mesh point using algorithms such as the Picard or the Newton iteration methods.

In the case of the chameleon EOM, we start by rescaling the equation (following our previous approach in Ref.~\cite{Briddon2021}): 

\begin{equation}
\alpha \hat{\nabla}^{2} \hat{\phi}=-\hat{\phi}^{-(n+1)}+\hat{\rho},
\label{eq:rescaled_EOM}
\end{equation}

\noindent where $\hat{\phi} = \phi/\phi_{\infty}$, $\hat{\rho} = \rho/\rho_{\infty}$, $\hat{\nabla}^{2} = L^{2} \nabla^{2}$, $L$ is the domain size and all the constants have been absorbed into the $\alpha$ parameter. The $\rho_{\infty}$ and $\phi_{\infty}$ here refer to the background density and the corresponding background value of the field. The $\alpha$ parameter is given by: 

\begin{equation}
\alpha \equiv\left(\frac{M \Lambda}{L^{2} \rho_{\infty}}\right)\left(\frac{n M \Lambda^{3}}{\rho_{\infty}}\right)^{\frac{1}{n+1}}.
\label{eq:alpha}
\end{equation}

To linearise Eq.~\ref{eq:rescaled_EOM} we can Taylor expand the non-linear term around some estimate of the field, $\phi_{k}$:

\begin{equation}
\begin{gathered}
\hat{\phi}^{-(n+1)} \approx \hat{\phi}_{k}^{-(n+1)}-(n+1) \hat{\phi}_{k}^{-(n+2)}\left(\hat{\phi}-\hat{\phi}_{k}\right)+\mathcal{O}\left(\hat{\phi}-\hat{\phi}_{k}\right)^{2} \\
\approx(n+2) \hat{\phi}_{k}^{-(n+1)}-(n+1) \hat{\phi}_{k}^{-(n+2)} \hat{\phi}+\mathcal{O}\left(\hat{\phi}-\hat{\phi}_{k}\right)^{2}.
\end{gathered}
\label{eq:taylor_expansion}
\end{equation}

\noindent We can then obtain the required variational (integral) form for the EOM, Eq.~\ref{eq:rescaled_EOM}, by applying Green's theorem and multiplying each term by $v_{j}$, which is an arbitrary test function that vanishes on the boundary of our domain, $\partial \Omega$. Finally, substituting the expansion for the non-linear term, Eq.~\ref{eq:taylor_expansion}, into the variational form of the EOM, we obtain:

\begin{equation}
\alpha \int_{\Omega} \hat{\nabla} \hat{\phi} \cdot \hat{\nabla} v_{j} \mathrm{~d} x+\int_{\Omega}(n+1) \hat{\phi}_{k}^{-(n+2)} \hat{\phi} v_{j} \mathrm{~d} x=\int_{\Omega}(n+2) \hat{\phi}_{k}^{-(n+1)} v_{j} \mathrm{~d} x-\int_{\Omega} \hat{\rho} v_{j} \mathrm{~d} x.
\label{eq:expanded_EOM}
\end{equation}

We can solve the non-linear EOM iterativly, by solving the above linearised version, Eq.~\ref{eq:expanded_EOM}, for some estimate of the field $\hat{\phi}_k$ and updating the estimate using


\begin{equation}
\hat{\phi}_{k+1}=\omega \hat{\phi}+(1-\omega) \hat{\phi}_{k},
\label{eq:picard_iteration}
\end{equation}

\noindent until the solution has converged to the desired degree of accuracy. The $\omega$ in Eq.~\ref{eq:picard_iteration} is the relaxation parameter, which controls the step size between the consecutive solutions found by the solver. This iterative procedure is called the Picard method.

To automate the solution to Eq.~\ref{eq:expanded_EOM}, we employ the SELCIE algorithm\footnote{SELCIE is available at: \url{https://github.com/C-Briddon/SELCIE}.} \cite{Briddon2021}. SELCIE is a FEM algorithm that automates the solution to the chameleon EOM for arbitrary density distributions. It allows for an easy mesh creation procedure as well as using different solvers to obtain solutions in 2D and 3D. SELCIE is based on the FEniCS Project software package, which is a collection of free and open-source software modules dedicated to automating solutions to differential equations via the FEM \cite{Logg2011, Logg2012, Alnaes2012, Scroggs2021}. Note that the approach laid out here is nearly identical to the methods previously applied to study chameleon gravity in NFW halos \cite{Tamosiunas2021}. The SELCIE software package and the procedure of solving the chameleon EOM is described in detail in Ref.~\cite{Briddon2021, Tamosiunas2021}. In summary, to obtain the solutions, we used the FEniCS in-built preconditioned Krylov solver with the conjugate gradient solver method and the default preconditioner settings (see section 4.3 in Ref.~\cite{Briddon2021} for further details). The mesh precision (the number of mesh cells per unit length) was set to 150. The code execution was set to stop when the absolute change in the field variable becomes smaller than $\delta \hat{\phi} = 10^{-14}$. The relaxation parameter, $\omega$, was set to 0.3, a value determined by experimentation.

\section{Cosmic Voids}
\label{cosmic_voids}

Cosmic voids are the most underdense regions in the Universe, found between the network of filaments in the cosmic web. The astrophysical properties of voids make them objects of special interest in cosmology and studies of gravity. Namely, being a major component of the cosmic web, as well as some of the largest objects in the Universe, voids hold a wealth of information about the underlying cosmology. The voids' structures, shapes and mutual alignments are sensitive to the properties of dark energy and gravity. Voids, being nearly empty (yet still populated by galaxies), also form a natural environment to study galaxy formation and evolution. Finally, the low densities found in voids make them a great environment for studying screening mechanisms.

The properties of cosmic voids have been studied extensively, both observationally and by using large scale structure formation simulations. Estimates of void sizes range between 10-50 $\mathrm{Mpc}/h$ \cite{Plionis2001, Hoyle2002}. A more recent study based on the Cosmic Void Catalog discovered similar results, with most void sizes falling between 10-30 $\mathrm{Mpc}/h$ \cite{Russell2017}. While the vast majority of the voids have sizes near the middle of the mentioned range, there are examples of extremely large/small voids. As an example, the Bootes void is know as one of the largest objects in the Universe with an approximate radial size of 44-50 $\mathrm{Mpc}/h$ \cite{Kirshner1987,Wegner2019}. On the lower side of the size spectrum, mini-voids have been detected with sizes in the range of 0.7-3.5 $\mathrm{Mpc}/h$ \cite{Karachentsev2004,Weygaert2011}.

There have been a number of studies investigating the density profiles of cosmic voids. Multiple universal void profiles have been proposed, based on both simulations and observational data. The simplest type of a profile explored in the literature is the step function (top hat) profile, described by:

\begin{equation}
\hat{\rho}_{\rm ST}(\hat{r})= \begin{cases}\hat{\rho}_{\rm out}, & \hat{r}>\hat{r}_{\rm step} \\ \hat{\rho}_{\rm in}, & \hat{r} \leq \hat{r}_{\rm step}\end{cases}    
\label{step_function_profile}.
\end{equation}

\noindent Here $\hat{\rho}_{\rm in}$ and $\hat{\rho}_{\rm out}$ refer to the inner and the outer density and $\hat{r}_{\rm step}$ is the radial position of the step. Note that the densities here are rescaled by the critical density of the Universe, while the radial variables are divided by the size of the domain of the problem at hand, $L$. While such a profile is simple to work with, and has been explored in the literature extensively, e.g. \cite{Clampitt2013}, in this work we will show that such a profile is not realistic enough to model the chameleon models that are left unconstrained by the current observational data. A more realistic profile choice is based on the hyperbolic tangent function, given by: 

\begin{equation}
    \hat{\rho}_{\rm th} (\hat{r}) = \frac{1}{2} \bigg[ (\hat{\rho}_{\rm in} + \hat{\rho}_{\rm out}) + (\hat{\rho}_{\rm out} - \hat{\rho}_{\rm in})\tanh(k(\hat{r} - \hat{r}_{\rm step}))    \bigg]. 
\label{tanh_profile}
\end{equation}

\noindent This profile converges to a Heaviside step function for $k \rightarrow \infty$. As before, the densities and the radial variables are rescaled.

In Ref.~\cite{Hamaus2014} a more physically motivated four-parameter profile based on observational data and simulation studies is introduced: 

\begin{equation}
 \hat{\rho}_{\mathrm{v}}(\hat{r}) -1=\delta_{\mathrm{c}} \frac{1-\left(\hat{r} / \hat{r}_{\mathrm{s}}\right)^{\alpha_{\mathrm{v}}}}{1+\left(\hat{r} / \hat{r}_{\mathrm{v}}\right)^{\beta_{\mathrm{v}}}}+\gamma_{\mathrm{v}}.
\label{void_profile}
\end{equation}

\noindent The $\delta_{\mathrm{c}}$ parameter here is the central density contrast, $\hat{r}_{\mathrm{v}}$ is the void radius rescaled by the domain size, $\hat{r}_{\mathrm{s}}$ is the rescaled scale radius at which $\rho_{\mathrm{v}} = \bar{\rho}$ (or $\hat{\rho}_{v} = 1$ in rescaled units) while $\alpha_{\mathrm{v}}$ and $\beta_{\mathrm{v}}$ determine the inner and outer slope of the void density distribution. The extra parameter $\gamma_{\mathrm{v}}$ is introduced in some works to account for the fact that a given void can be located in a under/over-dense region (e.g. as discussed in Ref.~\cite{Chantavat2017}). 

\begin{figure}
  \centering
    \includegraphics[width=0.60\textwidth]{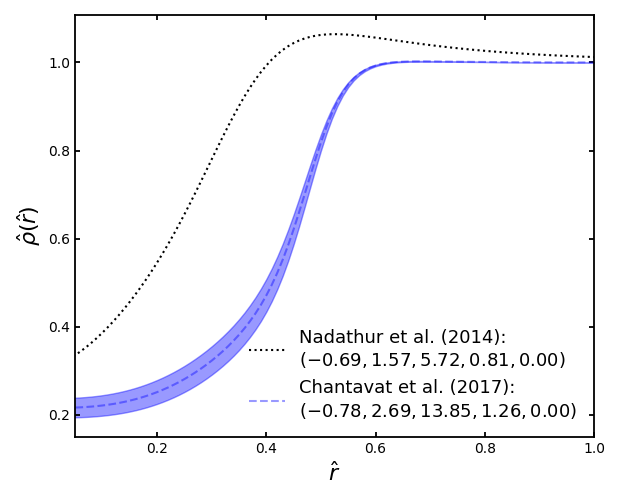}
    \caption{Void density profiles determined from the CMB and SDSS data and simulations as discussed in \cite{Chantavat2017} and \cite{Nadathur2014}. The density is rescaled by the value of the critical density of the Universe, while the radial value is divided by the size of the domain: $L = 2 \times 31.68 \; \mathrm{Mpc}/h$, i.e. twice the typical size of a void. The values in the brackets refer to the best-fit parameters: ($\delta_{\rm c}$, $\alpha_{\rm v}$, $\beta_{\rm v}$, $\hat{r}_{\rm s}/\hat{r}_{\rm v}$, $\gamma_{\rm v}$). The blue band corresponds to the standard deviation around the best-fit values.} 
    \label{void_density_profiles}
\end{figure}

\begin{figure}
\centering
\makebox[\linewidth][c]{%
  \begin{subfigure}[b]{0.4\textwidth}
  \centering
    \includegraphics[width=1.00\textwidth]{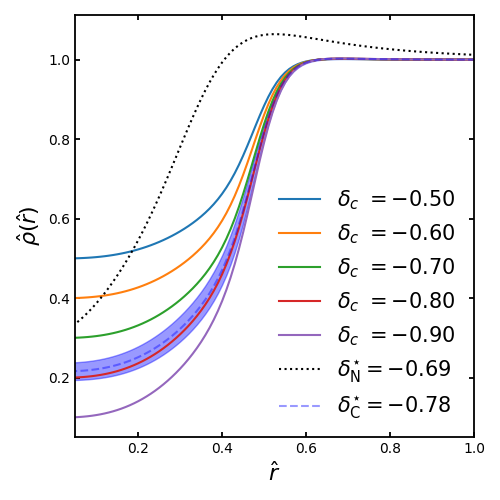}
    \label{density_deltas}
  \end{subfigure}
  \begin{subfigure}[b]{0.4\textwidth}
  \centering
    \includegraphics[width=1.00\textwidth]{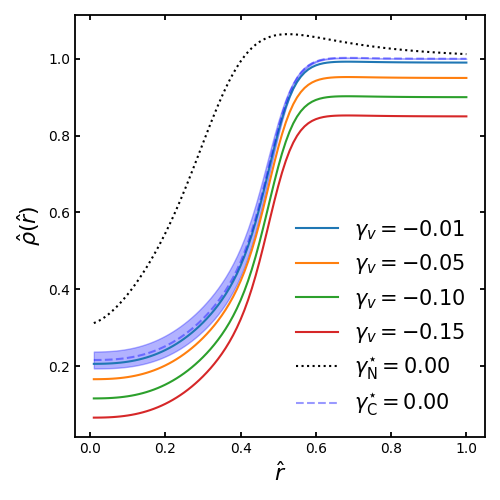}
    \label{density_gammas}
  \end{subfigure}
  }
  \caption{Void density profiles for different $\delta_{\rm c}$ (void depths) and $\gamma_{\rm v}$ parameters. As before, the other parameters are set to the best-fit values from \cite{Chantavat2017} and \cite{Nadathur2014}. Note that all quantities are rescaled as defined in section \ref{finite_element_method}.}
  \label{density_varying_params2}
\end{figure}

In Ref.~\cite{Nadathur2014, Nadathur2015} a variation of the profile in Eq.~\ref{void_profile} is fitted to simulated void data. Specifically, the void radius is replaced with the scale radius in the denominator of Eq.~\ref{void_profile} and fitted to realistic mock luminous red galaxy (LRG) catalogues from the Jubilee simulation, as well as void catalogues constructed from the SDSS LRG and Main Galaxy samples \cite{Nadathur_Hotchkiss2014, Watson2013, Nadathur2015}. The voids were detected using the watershed transform algorithm ZOBOV \cite{Neyrinck2008}. The self-similarity and universality of void density profiles were further studied in Ref.~\cite{Ricciardelli2014}, while the velocity profiles were studied in Ref.~\cite{Massara2018}. In Ref.~\cite{Chantavat2017} the profile described in Eq.~\ref{void_profile} is fitted to the lensing potential map from Planck CMB lensing reconstruction analysis data in combination with the “Public Cosmic Void Catalog”. The analysis in Ref.~\cite{Chantavat2017} determined that introducing an extra parameter, $\gamma_{\mathrm{v}}$, improves the fit to the data. This is done to account for the fact that Sloan Digital Sky Survey voids reside in an underdense region. In our analysis we will employ the profiles and the best-fit numbers from Ref.~\cite{Nadathur2014, Chantavat2017}
as a benchmark to compare the calculated results against. Specifically, for Ref.~\cite{Nadathur2014}, the following best-fit numbers will be used: $\delta_{\mathrm{c}}=-0.69$, $\hat{r}_{\mathrm{s}}=0.81 \hat{r}_{\mathrm{v}}$, $\alpha_{\mathrm{v}} = 1.57$, $\beta_{\mathrm{v}} = 5.72$. Similarly, for the best-fit profile from Ref.~\cite{Chantavat2017}, we will use $\delta_{\mathrm{c}}=-0.78$, $\hat{r}_{\mathrm{s}}=1.26 \hat{r}_{\mathrm{v}}$, $\alpha_{\mathrm{v}} = 2.69$, $\beta_{\mathrm{v}} = 13.85$. For the void size we will use a value of $r_{\mathrm{v}} = 31.68$ Mpc/$h$, which is a good estimate for the size of a typical void as discussed in Ref \cite{Chantavat2017}. The $\gamma_{\mathrm{v}}$ parameter will generally be set to zero, but, for the sake of completeness, the effects of the mentioned parameter will be investigated in Fig.~\ref{density_varying_params2}. The referenced profiles are plotted in Fig.~\ref{void_density_profiles}, where we see that there is significant variation between the best-fit void profiles in different works in the literature. This is not entirely surprising, as the best-fit parameters in Ref.~\cite{Nadathur2014, Chantavat2017} have been determined using different techniques and come with different systematics and errors. In this work this will not be an issue, as we will only use the outlined profiles as a benchmark, and the void parameters will be varied to explore a large set of possible profiles.

Given the form of Eq.~\ref{void_profile}, one can derive an analytical expression for the void mass enclosed by some radius $R$ assuming spherical symmetry: 

\begin{equation}
\begin{split}
    M_{\mathrm{v}}(R) = \int_{0}^{R} 4 \pi r^{2} \rho_{\mathrm{v}}(r) dr &= 4 \pi \rho_{c} R^{3} \Bigg[ \frac{1}{3} + \frac{\delta_{c}}{3(\alpha_{\mathrm{v}}+3)} \bigg((\alpha_{\mathrm{v}}+3) \; { }_{2} F_{1}\bigg(1, \frac{3}{\beta_{\mathrm{v}}} ; \frac{\beta_{\mathrm{v}}+ 3}{\beta_{\mathrm{v}}} ; - \bigg(\frac{R}{r_{\mathrm{v}}}\bigg)^{\beta_{\mathrm{v}}} \bigg) \\  
    &\quad - 3\bigg(\frac{R}{r_{\mathrm{s}}}\bigg)^{\alpha_{\mathrm{v}}} { }_{2} F_{1}\bigg(1, \frac{\alpha_{\mathrm{v}}+3}{\beta_{\mathrm{v}}}; \frac{\alpha_{\mathrm{v}} + \beta_{\mathrm{v}} + 3 }{\beta_{\mathrm{v}}}; \bigg( \frac{R}{r_{\mathrm{s}}} \bigg)^{\beta_{\mathrm{v}}}  \bigg) \bigg) + \frac{\gamma_{\mathrm{v}}}{3} \Bigg], 
\end{split}
\label{void_mass}
\end{equation}

\noindent where ${ }_{2} F_{1}$ corresponds to the hypergeometric function, defined as: 

\begin{equation}
{ }_{2} F_{1}(a, b ; c ; z)=\sum_{n=0}^{\infty} \frac{(a)_{n}(b)_{n}}{(c)_{n}} \frac{z^{n}}{n !}=1+\frac{a b}{c} \frac{z}{1 !}+\frac{a(a+1) b(b+1)}{c(c+1)} \frac{z^{2}}{2 !}+\cdots
\label{hypergeometric_function}
\end{equation}

\noindent Integrating the Poisson equation then gives the relation for the regular Newtonian gravitational acceleration:

\begin{equation}
a_{\mathrm{Newt}}(R) = -\frac{GM_{\mathrm{v}}(R)}{R^{2}}.
\label{Newtonian_acceleration}    
\end{equation}

\section{Analytic Investigation}
\label{analytic_investigation}

When it comes to analyzing the effects of cosmic void density profiles on chameleon screening, a lot of headway can be made analytically. As previously discussed in Ref.~\cite{Tamosiunas2021}, for much of the allowed parameter space (when dealing with cosmological length scales) the chameleon field tracks the minimum value of its effective potential.
More specifically, in terms of the EOM (Eq.~\ref{eq:EOM}), this refers to the regime where $\alpha \nabla^2 \phi \ll V'_{\rm eff}(\phi)$, with $V_{\rm eff}(\phi) = V(\phi) + \rho \phi/M$. We will refer to this as the small-$\alpha$ regime. In this regime we can use the approximation: 

\begin{equation}
\hat{\phi}(\hat{r}) \approx [\hat{\rho}_{\mathrm{v}}(\hat{r})]^{-\frac{1}{n+1}}.
\label{small_alpha_approx}
\end{equation}

\noindent An approximate analytic expression for the field gradient is therefore given by the derivative of Eq.~\ref{small_alpha_approx}:

\begin{equation}
    \frac{\partial \hat{\phi}(\hat{r})}{\partial \hat{r}} \approx  -\frac{[\hat{\rho}_{\rm v}(\hat{r})]^{-\frac{n+2}{n+1}}}{n + 1} \frac{\partial \hat{\rho}_{\rm v}(\hat{r})}{\partial \hat{r}}.
\label{void_grad_phi_analytic}
\end{equation}

\begin{figure}
\centering
\makebox[\linewidth][c]{%
  \begin{subfigure}[b]{0.37\textwidth}
  \centering
    \includegraphics[width=1.00\textwidth]{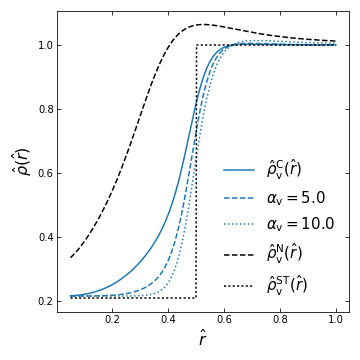}
    \label{analytic_results_rho}
  \end{subfigure}
  \begin{subfigure}[b]{0.37\textwidth}
  \centering
    \includegraphics[width=1.00\textwidth]{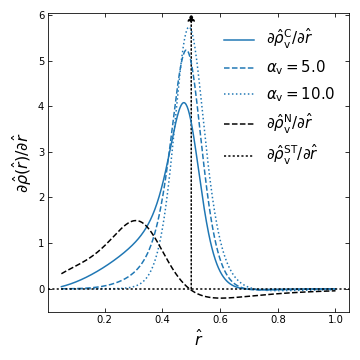}
    \label{analytic_results_grad_rho}
  \end{subfigure}
  \begin{subfigure}[b]{0.37\textwidth}
  \centering
      \includegraphics[width=1.00\textwidth]{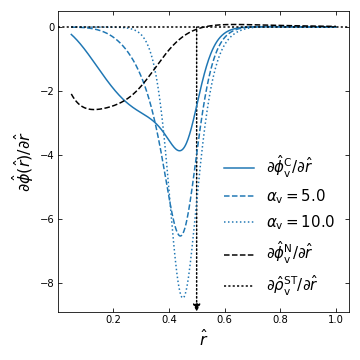}
      \label{analytic_results_grad_phi}
  \end{subfigure}
  }
  \caption{The results of the analytic investigation. \textbf{Left:} different density profiles. The black dashed and the blue solid lines correspond to the profiles described in Fig.~\ref{void_density_profiles}. The dashed and the dotted blue lines were generated by varying the $\alpha_{\rm v}$ parameter value, while keeping the other parameters fixed to the best-fit values used for the solid blue line (i.e. the Ref.~\cite{Chantavat2017} values). The dotted line corresponds to the step function density profile. The arrow corresponds to the Dirac delta function (i.e. the value goes off to infinity). \textbf{Centre:} the corresponding radial derivative of the different density profiles. \textbf{Right:} the corresponding radial gradients of the chameleon field. Note that all quantities are rescaled as defined in section \ref{finite_element_method}.} 
  \label{analytic_results}
\end{figure}

\noindent Eq.~\ref{void_grad_phi_analytic} illustrates that the chameleon field depends strongly on the shape of the void density profile. This behaviour is illustrated in Fig.~\ref{analytic_results}. In particular, we plot different density profiles, starting with the universal density profile described in Fig.~\ref{void_density_profiles} and ending with a step function (top hat) profile. For the more realistic Hamaus-Chantavat-type profile (Eq.~\ref{void_profile}), we also vary the $\alpha_{\mathrm{v}}$ parameter (which controls the shape of the inner slope of the void), while keeping the other parameters fixed to their best-fit values. A general conclusion that can be made here is that the more step function-like a profile is, the larger the corresponding radial derivative of the chameleon field. The radial derivative of the field is maximised for the step function profile in the region of the step. In this case, the radial derivative of the density and the corresponding chameleon field goes to infinity. For this reason, the analytic approximation for the small-$\alpha$ regime, Eq.~\ref{small_alpha_approx}, cannot be used to accurately describe the behaviour of the field at the step region. The full numerical solutions that we will describe in Section \ref{numerical_investigation} are required to compute the chameleon field at the step region. 

\begin{figure}
\centering
\makebox[\linewidth][c]{%
  \begin{subfigure}[b]{0.37\textwidth}
  \centering
    \includegraphics[width=1.00\textwidth]{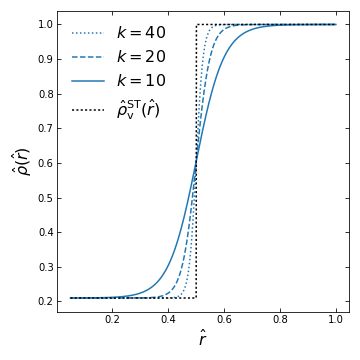}
    \label{analytic_results_tanh_rhos}
  \end{subfigure}
  \begin{subfigure}[b]{0.37\textwidth}
  \centering
    \includegraphics[width=1.00\textwidth]{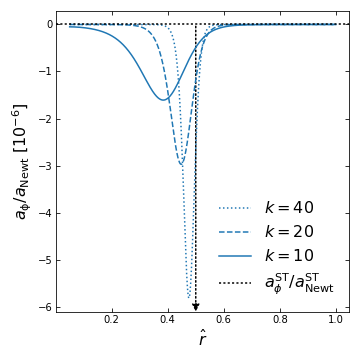}
    \label{acceleration_ratios_tanh}
  \end{subfigure}
  \begin{subfigure}[b]{0.37\textwidth}
  \centering
      \includegraphics[width=1.00\textwidth]{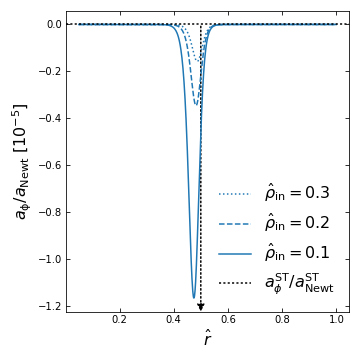}
      \label{acceleration_ratios_tanh_depths}
  \end{subfigure}
  }
  \caption{The results of the analytic investigation for a tanh density profile. \textbf{Left:} different density profiles for different values of the $k$ parameter. \textbf{Center:} the corresponding chameleon-to-Newtonian acceleration ratios. \textbf{Right:} acceleration ratios for different inner void densities (i.e. different depths). In this case each profile is calculated with $k = 40$ and $\hat{\rho}_{\mathrm{out}} = 1.0$. In all cases the dotted black line corresponds to the step function profile, while the black arrow denotes the profile going infinite (Dirac delta function). Note that all quantities are rescaled as defined in section \ref{finite_element_method}.}
  \label{analytic_results_tanh_rho}
\end{figure}

While the step function profile is clearly not a realistic description of the density distribution inside of a void, it is an interesting limiting case. To investigate this limiting case further we will explore the hyperbolic tangent profile described in Eq.~\ref{tanh_profile}. Such a profile converges to a step function for large values of $k$. The resulting ratios between the chameleon and the Newtonian acceleration are shown in Fig.~\ref{analytic_results_tanh_rho}. Specifically, the acceleration ratios are shown for different values of $k$ along with the different depths of the void. The key finding here is that the acceleration ratios are primarily limited by the inner slope of the void profile along with the inner depth of the density distribution. In principle one could obtain arbitrarily large values for the acceleration ratio by increasing the inner slope of the density distribution. As before, however, for large values of $k$ the small-$\alpha$ approximation breaks down and full numerical treatment is needed. By applying a small perturbation to the field value given by Eq.~\ref{small_alpha_approx}, we can find that the small-$\alpha$ approximation will hold only if $\sqrt{\alpha} \ll 1/k$. For $\alpha$ to lie in the allowed regions of the parameter space it must satisfy $\alpha < 10^{-12}$. Therefore, so long as $k \ll 10^6$, the small-$\alpha$ approximation will hold for all values of $\alpha$ allowed by the parameter space.

\section{Numerical Investigation}
\label{numerical_investigation}

\begin{figure}
\centering
\makebox[\linewidth][c]{%
  \begin{subfigure}[b]{0.37\textwidth}
  \centering
    \includegraphics[width=1.00\textwidth]{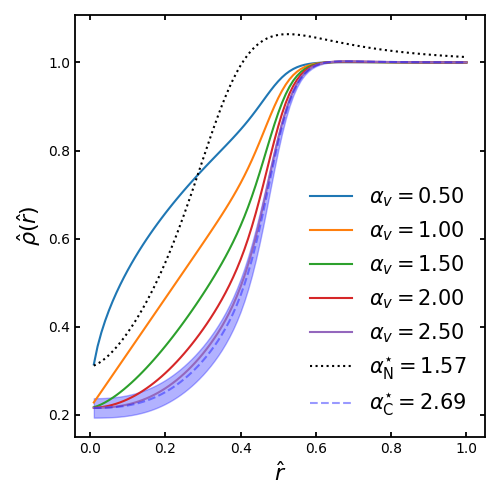}
    \label{density_alphas}
  \end{subfigure}
  \begin{subfigure}[b]{0.37\textwidth}
  \centering
    \includegraphics[width=1.00\textwidth]{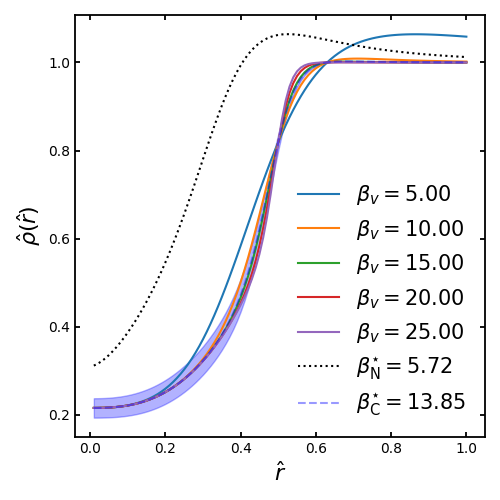}
    \label{density_betas}
  \end{subfigure}
  \begin{subfigure}[b]{0.37\textwidth}
  \centering
      \includegraphics[width=1.00\textwidth]{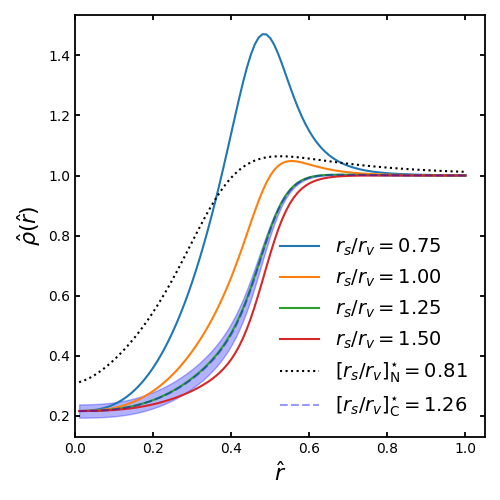}
      \label{density_rs_rv}
  \end{subfigure}
  }
  \caption{Void density profiles for different free parameters. The $\alpha_{\mathrm{v}}$, $\beta_{\mathrm{v}}$ and $\hat{r}_{s}/\hat{r}_{\mathrm{v}}$ parameters are varied, while keeping the other parameters fixed to their best-fit values in each case (i.e. we are varying the best-fit parameter values of the blue profile in Fig.~\ref{void_density_profiles}). As before the black dotted line and the blue band refers to the best-fit profiles from Ref. \cite{Chantavat2017} and \cite{Nadathur2014}. Note that all quantities are rescaled as defined in section \ref{finite_element_method}.}
  \label{density_varying_params1}
\end{figure}

\begin{figure}
\centering
\makebox[\linewidth][c]{%
  \begin{subfigure}[b]{0.365\textwidth}
  \centering
    \includegraphics[width=1.00\textwidth]{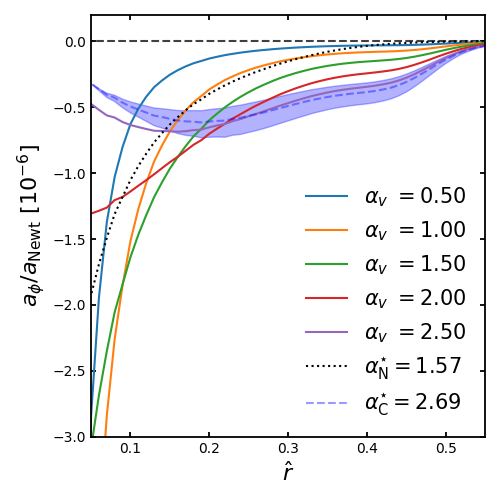}
    \label{acceleration_alphas}
  \end{subfigure}
  \begin{subfigure}[b]{0.37\textwidth}
  \centering
    \includegraphics[width=1.00\textwidth]{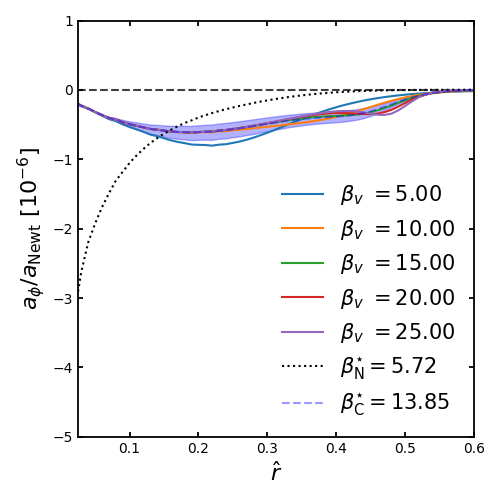}
    \label{acceleration_betas}
  \end{subfigure}
  \begin{subfigure}[b]{0.37\textwidth}
  \centering
      \includegraphics[width=1.00\textwidth]{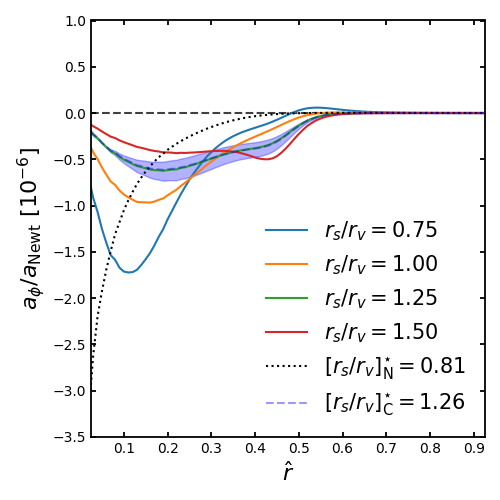}
      \label{acceleration_rs_rv}
  \end{subfigure}
  }
  \caption{The chameleon-to-Newtonian acceleration ratios for density profiles depicted in Fig.~\ref{density_varying_params1}. }
  \label{acceleration_varying_params1}
\end{figure}

\begin{figure}
\centering
\makebox[\linewidth][c]{%
  \begin{subfigure}[b]{0.4\textwidth}
  \centering
    \includegraphics[width=1.00\textwidth]{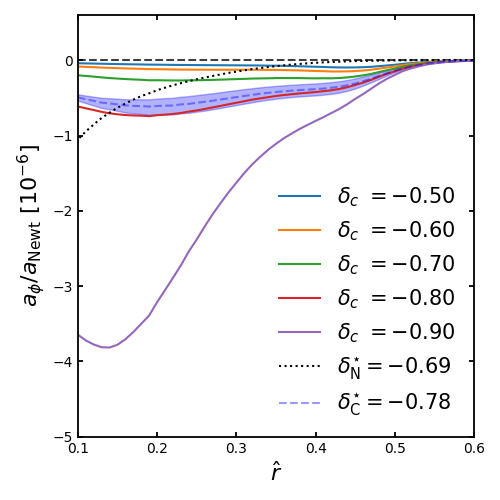}
    \label{acc_ratio_deltas}
  \end{subfigure}
  \begin{subfigure}[b]{0.4\textwidth}
  \centering
    \includegraphics[width=1.00\textwidth]{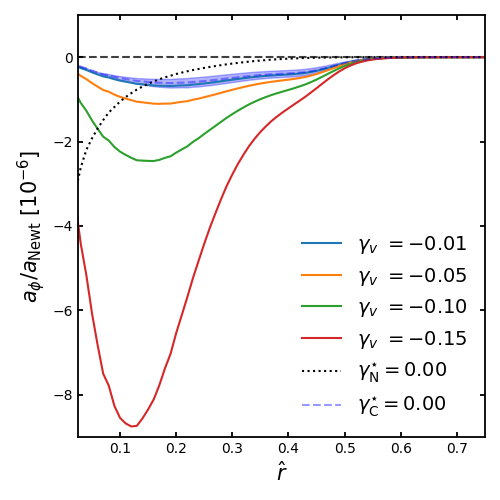}
    \label{acc_ratio_gammas}
  \end{subfigure}
  }
  \caption{Chameleon-to-Newtonian acceleration ratios corresponding to the density profiles depicted in Fig.~\ref{density_varying_params2}.}
  \label{acceleration_varying_params2}
\end{figure}

In this section we turn our attention to more realistic density distributions. In particular, we explore the Nadathur and Hamaus-Chantavat-type density profiles from Ref.~\cite{Nadathur2014, Chantavat2017} (as depicted in Fig.~\ref{void_density_profiles}). Each free parameter is varied (while keeping the other parameters fixed to their best-fit values). For each set of the parameters we solve the EOM numerically by employing SELCIE, as described in more detail in section \ref{finite_element_method}. Fig.~\ref{density_varying_params1} depicts the various density profiles generated by varying the $\alpha_{\mathrm{v}}$, $\beta_{\mathrm{v}}$ and the $\hat{r}_{\mathrm{s}}/\hat{r}_{\mathrm{v}}$ parameters, while Fig.~\ref{acceleration_varying_params1} shows the corresponding chameleon-to-Newtonian acceleration ratios. Fig.~\ref{density_varying_params1} and \ref{acceleration_varying_params2} show the corresponding results for voids of varying depths. 

We find that for the more realistic profiles, as depicted in Fig.~\ref{density_varying_params1}, the acceleration ratio values are generally lower than in the case of the step function and the tanh density profiles, Eq.~\ref{tanh_profile}. A key finding is that the parameters that control the depth of the void along with the inner density slope (i.e. $\delta_{\mathrm{c}}$ and $\alpha_{\mathrm{v}}$) have the most significant effect on the values of the acceleration ratios. Similarly, the $\gamma_{\mathrm{v}}$ parameter has a similar effect to the $\delta_{\mathrm{c}}$ parameter through its effect on the central depth of the void.

An interesting observation that can be drawn from the $\alpha_{\mathrm{v}}$ variation plot (Fig.~\ref{acceleration_varying_params1}) is that there are generally two regimes of behaviour. Specifically, for $\alpha_{\mathrm{v}} \lesssim 2$ the acceleration ratios reach maximal values towards the centre of the void. However, for $\alpha_{\mathrm{v}} \gtrsim 2$, the acceleration ratio curve falls to some maximally negative value before turning back towards zero in the central region of the void. Such behaviour is determined by the gradient of the density profile in the central region of the void along with the slope closer to the region where the void density reaches its maximum value, i.e. the compensation wall. On the other hand, varying $\beta_{\mathrm{v}}$ and the $\hat{r}_{\mathrm{s}}/\hat{r}_{\mathrm{v}}$ ratio only changes the slope in the outer regions of the void, thus the two regimes of behaviour are not observed in those cases. The depth of the void is also of key importance for determining the maximum value of the acceleration ratio. Generally, the deeper the void, the larger the chameleon-to-Newtonian acceleration ratio. This can be achieved by either having a small value of $\delta_{\mathrm{c}}$ or a large negative value of $\gamma_{\mathrm{v}}$. 

\section{Conclusions}

In this work we have investigated the surviving chameleon models in the context of cosmic voids. Specifically, we have studied chameleon models still allowed by the observational and laboratory constraints and solved the chameleon EOM given a selection of void density profiles using both analytic and numerical techniques. In the case of a step function profile, we showed that the chameleon acceleration vanishes in all regions except for the step itself. At the step, the radial derivative of the density profile goes to infinity, which, under the small-$\alpha$ approximation, would result in an infinite chameleon acceleration. However, as discussed in section \ref{analytic_investigation}, the small-$\alpha$ approximation is not valid in profiles of sufficient steepness, hence full numerical treatment is required, which results in large (but finite) values at the step region. In summary, while the step-function profile is not physically realistic, it is nonetheless useful as it acts as a limiting case that maximises the chameleon acceleration. In other words, the more step function-like a void is, the higher is the corresponding chameleon acceleration. While, in principle, the void acceleration is only limited by the steepness of the inner density gradient, in practice, for voids of realistic gradients, chameleon-to-Newtonian acceleration ratios of $a_{\phi}/a_{\rm Newt} \sim 10^{-6} - 10^{-5}$ can be obtained for the parameter values studied in this work.  

When investigating the more physically motivated profiles we have shown that the chameleon field profiles exhibit a rich and complex behaviour. In summary, the chameleon-to-Newtonian acceleration ratios are most sensitive to the inner density slope and the depth of the void. The variation of the inner density slope, in particular, can lead to two different behaviours -- the acceleration ratios reaching maximum (negative) values towards the centre of the void or somewhere between the center and the compensation wall of the void. However, irrespective of which of these two regimes we consider, the acceleration ratios reach similar values of $\sim 10^{-6}$.

It should be noted that chameleon models are not limited to the scalar-tensor theories studied in this work. In particular, a subset of $f(R)$ theories, such as Hu-Sawicki gravity, also exhibit chameleon screening \cite{Hu-Sawicki2007}. It can be shown that $f(R)$ theories are equivalent to scalar-tensor theories with a particular choice for the potential (e.g. see Ref.~\cite{Velasquez2018}). The Hu-Sawicki theory, in particular, corresponds to different scalar-tensor parameter values, which lead to acceleration ratios that are much higher. The approximate acceleration ratios for the Hu-Sawicki model are calculated in appendix \ref{appendix A}. Currently with our code we can only calculate approximate results for the Hu-Sawicki model. This is the case, as the code cannot handle the negative fractional values for the $n$ parameter. To obtain a direct comparison, a new implementation of the code would have to be written, with a specific FEM approach applied to solve the $f(R)$ EOM directly. This is left for future work.  

Our results shine light on the observational prospects of detecting the effects of the chameleon field in cosmic voids. For the parameter space region studied in this work, the acceleration ratios of $\approx 10^{-6}-10^{-5}$ would be extremely challenging to detect. Nonetheless, even such a small fifth force could potentially be detected statistically through its effects on structure formation. A fifth force, in the central region of a void, or near the compensation wall, could potentially increase the average void depth and the size of the typical compensation wall. To investigate such effects in detail, large scale structure simulations with modified gravity effects would be required. Ideally, one would run hydrodynamic modified gravity simulations to investigate the possible degeneracies between astrophysical effects and effects due to a fifth force. Other models, such as the mentioned Hu-Sawicki model are known to exhibit much larger fifth forces, which have been and will be further constrained by previous and upcoming observational data (e.g. galaxy cluster weak lensing measurements) \cite{Wilcox2015,Pizzuti2017}.

The study described in this work can be extended in multiple ways. For instance, here we had made a key assumption of voids being spherically symmetric. Real cosmic voids are better described as triaxial objects. While generally voids tend to become more spherical with expansion, their axis ratios vary between $\sim 0.5-1.0$ \cite{Weygaert2011}. Hence, analogously to NFW halos, which we had previously studied in Ref.~\cite{Tamosiunas2021}, we expect fifth force effects that depend on the void shape. Nonetheless, we do not expect these effects to significantly increase or reduce the chameleon-to-Newtonian acceleration ratios. Effects due to void shapes could also be investigated in detail by employing modified gravity simulations. 

Finally, it should be noted that in this work we had only investigated simple void systems -- i.e. single voids of different depths. There are, however, other more complicated systems that could be investigated numerically. As an example, a system of multiple voids separated by filaments would lead to a large gradient in the density distribution, which, in turn, would lead to relatively high values of the chameleon accleration. Such systems could be investigated with SELCIE; however, a more detailed study would require hydrodynamical modified gravity simulations. On a similar note, such systems are also of special interest in other modified gravity models, e.g. the symmetron, which are currently less constrained. The symmetron-type models exhibit complex time-dependent behaviour (e.g. domain walls). Such behaviour could be investigated by employing the SELCIE code; however, at the time of writing, the code does not allow for the solving of time-dependent equations. We are currently working towards expanding the code functionality to include a solver for the symmetron EOM and to model time-dependent systems.

\appendix
\section{Acceleration Ratios for Gravitational Strength Couplings}
\label{appendix A}

The acceleration ratios obtained in this work are significantly smaller that those found in the literature (e.g. Ref.~\cite{Shao2019, Wilson2021}), where the fifth force is found to be of similar magnitude to the Newtonian force inside cosmic voids. The key reason for this discrepancy in results is due to the fact that the mentioned works consider a different model, i.e. the $f(R)$ Hu-Sawicki model \cite{Hu-Sawicki2007}. The Hu-Sawicki model can be mapped directly to the parameter space of the scalar-tensor chameleon models explored in this work. Namely, the Hu-Sawicki model corresponds to a different part of the $\{n, M, \Lambda\}$ parameter space, resulting in significantly higher fifth force. In the rest of this section, we lay out the mapping between the scalar-tensor models described in this work and the corresponding $f(R)$ theories. Furthermore, we calculate the acceleration ratios for a particular case of $f(R)$ gravity, which is comparable to the Hu-Sawicki model.  

The Hu-Sawicki model refers to a specific case of $f(R)$ gravity with:

\begin{equation}
f(R)=-m^2 \frac{c_1\left(R / m^2\right)^{n_{1}}}{c_2\left(R / m^2\right)^{n_{1}}+1}
\label{Hu-sawicki},
\end{equation}

\noindent where $m, n_{1}, c_{1}$ and $c_{2}$ are the model parameters. Our scalar-tensor model can be directly compared to $f(R)$ by noting that: 

\begin{equation}
f_R=-\sqrt{\frac{2}{3}} \frac{\phi}{M_{\mathrm{Pl}}},
\label{f_R_equation}
\end{equation}

\noindent with $f_{R} \equiv \partial f/\partial R$ and $\phi$ as the chameleon scalar field. For the current value of the $f_{R}$ parameter we have: 

\begin{equation}
f_{R 0}=-\sqrt{\frac{2}{3}} \frac{\phi_{\infty}}{M_{\mathrm{Pl}}},
\label{fR0}
\end{equation}

\noindent with $\phi_{\infty}$ as the field value at $r \rightarrow \infty$. Explicitly, the background value of the field is: 

\begin{equation}
\phi_{\infty}=\left(\frac{n M \Lambda^{n+4}}{\rho_{\infty}}\right)^{\frac{1}{n+1}},
\label{phi_infinity}
\end{equation}

\noindent where $\rho_{\infty}$ is the corresponding density. Substituting the field value into Eq. \ref{fR0} gives: 

\begin{equation}
\left|f_{R 0}\right|=\sqrt{\frac{2}{3}} \frac{1}{M_{\mathrm{Pl}}}\left[\frac{n M \Lambda^{n+4}}{\rho_{\infty}}\right]^{\frac{1}{n+1}}.
\end{equation}

\noindent Now we should note that the coupling parameter in $f(R)$ gravity is set to $\beta=\sqrt{1 / 6}$ or equivalently $M =\sqrt{6} M_{\mathrm{Pl}}$. Furthermore, we should note that models with $|f_{R0}| = \{10^{-4}, 10^{-5}, 10^{-6} \}$ are often considered in the literature. Finally, if we set $n = 1$\footnote{Here we should note that the choice of $n = 1$ does not directly correspond to the Hu-Sawicki model. For a direct comparison we would need to set $n$ to a negative fractional value, which currently our code cannot handle. Nonetheless, after experimenting with different values, we found that the results do not strongly depend on $n$. Hence our results should be a good approximation of the acceleration ratios in the Hu-Sawicki model.} for convenience and $\rho_{\infty} = \rho_{c}$ we can then solve for $\Lambda$, i.e. the $\Lambda$ value that corresponds to the outlined $f_{R0}$ values in the corresponding $f(R)$ model:

\begin{equation}
\Lambda=\frac{3 \rho_{\mathrm{c}} M_{\mathrm{Pl}}\left|f_{R 0}\right|^2}{\sqrt{24}}.
\label{Lambda_HS}
\end{equation}

\begin{figure}
\centering
\makebox[\linewidth][c]{%
  \begin{subfigure}[b]{0.4\textwidth}
  \centering
    \includegraphics[width=1.00\textwidth]{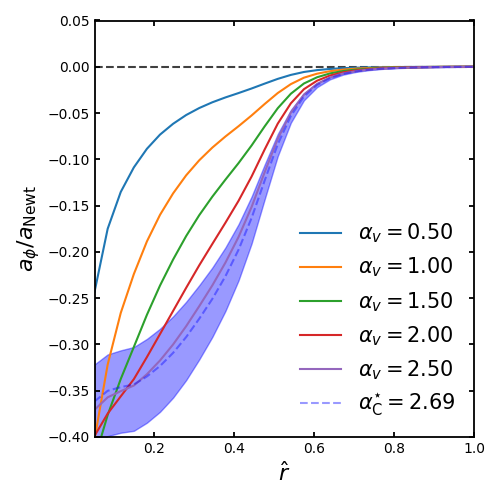}
    \label{acc_ratios_f(R)_alphas}
  \end{subfigure}
  \begin{subfigure}[b]{0.395\textwidth}
  \centering
    \includegraphics[width=1.00\textwidth]{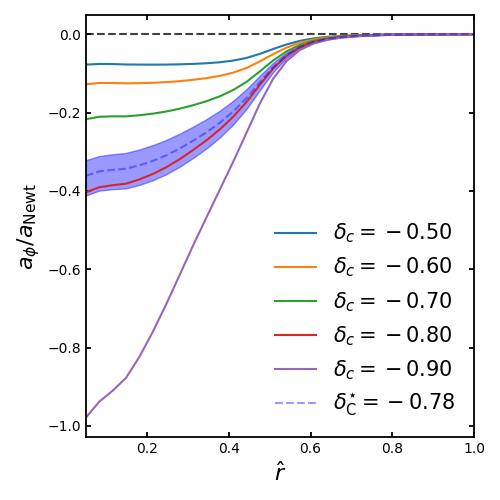}
    \label{acc_ratios_f(R)_deltas}
  \end{subfigure}
  }
  \caption{The acceleration ratios for $f(R)$ gravity with $|f_{R0}| = 10^{-5}$. The ratios were calculated using the best-fit parameter values for the density profiles from Ref.~\cite{Chantavat2017} as shown in Fig.~\ref{void_density_profiles}, with the $\alpha_{\mathrm{v}}$ and $\delta_{\mathrm{c}}$ parameters varied correspondingly.}
  \label{acceleration_ratios_f(R)}
\end{figure}

\noindent Numerically, if we plug in $\rho_{\rm c} = 8.07 \times 10^{-11} h^{2}$ $\mathrm{eV^{4}}$, $|f_{R0}| = 10^{-5}$ and $M_{\rm Pl} = 2.435 \times 10^{27}$ eV, we get $\Lambda = 22.6$ eV. Finally, if we plug in the $n$, $M$ and $\Lambda$ values outlined above to our numerical code, it allows us to compute the approximate values for the acceleration ratios in Hu-Sawicki gravity. For this we used the Chantavat et al. (2017) best-fit numbers (Fig.~\ref{void_density_profiles}) and varied the $\alpha_{\rm v}$ and $\delta_{\rm c}$ parameters. This results in the acceleration ratios given in Fig.~\ref{acceleration_ratios_f(R)}. The obtained values are significantly higher than those reported in the previous figures. In particular, the acceleration ratios generally agree with the previous results in the literature (e.g. Ref.~\cite{Wilson2021}). Specifically, Table II in Ref.~\cite{Wilson2021} lists the acceleration ratio of $0.19 \pm 0.002$ for void sizes in the range of $25-35 \; \mathrm{Mpc}/h$, redshift $z = 0.5$ and $|f_{R0}| = 10^{-5}$. Meanwhile, our results vary between $|a_{\phi}/a_{\rm Newt}| \approx 0.1 - 1.0$ depending on the density profile used. Here it should be noted that minor differences between the results are to be expected as the acceleration values were calculated using different datasets, corresponding to voids at different redshifts, of different sizes and different density profiles.

\acknowledgments

We would like to thank Seshadri Nadathur for his insights into different void density profiles and various void finding algorithms. 

Clare Burrage and Andrius Tamosiunas are supported by a Research Leadership Award from The Leverhulme Trust. Chad Briddon is supported by the University of Nottingham. Adam Moss is supported by a Royal Society University Research Fellowship.


\clearpage
\bibliography{refs} 
\bibliographystyle{JHEP}

\end{document}